\documentclass{webofc}
\usepackage[varg]{txfonts}   
\usepackage{listings}
\usepackage{hyperref}
\begin{document}
\title{Synergies between Exoplanet Surveys and Variable Star Research}
%
%

\author{\firstname{Geza} \lastname{Kovacs}\inst{1}\fnsep\thanks{\href{mailto:kovacs@konkoly.hu}{\tt kovacs@konkoly.hu}} 
}

\institute{Konkoly Observatory of the Hungarian Academy of Sciences, \\ 
H-1121, Budapest, Konkoly Thege M. ut 15-17 
          }

%
%
\abstract{
With the discovery of the first transiting extrasolar planetary system 
back to 1999, a great number of projects started to hunt for other similar 
systems. Because of the incidence rate of such systems was unknown and the 
length of the shallow transit events is only a few percent of the orbital 
period, the goal was to monitor continuously as many stars as possible for 
at least a period of a few months. Small aperture, large field of view 
automated telescope systems have been installed with a parallel development 
of new data reduction and analysis methods, leading to better than 1\% per 
data point precision for thousands of stars. With the successful launch of 
the photometric satellites CoRot and Kepler, the precision increased further 
by one-two orders of magnitude. Millions of stars have been analyzed and 
searched for transits. In the history of variable star astronomy this is the 
biggest undertaking so far, resulting in photometric time series inventories 
immensely valuable for the whole field. In this review we briefly discuss 
the methods of data analysis that were inspired by the main science driver 
of these surveys and highlight some of the most interesting variable star 
results that impact the field of variable star astronomy. 
}
\maketitle
%

%
%
\section{Introduction - some historical background}\label{sec:sec-1}
Up until 1999, telephoto lenses were not acknowledged as the source of major 
astronomical discoveries. This has changed dramatically when, after the discovery 
of the first extrasolar planet 51~Peg \cite{mayor1995}, \cite{marcy1995}, many 
research teams started concentrated efforts to observe the photometric transit 
signals of the already discovered planets, and find new ones. Because some of 
the systems discovered by radial velocity methods had short periods of a few days 
and the estimated masses were in the range of Jupiter masses, these extrasolar 
planets (dubbed as Hot Jupiters, or HJs for short) with their $5-20$\% 
transit probabilities and $\sim1$\% transit depths were excellent targets for 
photometric detection, even for telescopes with apertures of $\sim 5-10$~cm. 
Indeed, five years after the discovery of 51~Peg, the photometric monitoring of 
HD~209458 resulted in the multiple observations of the transit events of the 
companion \cite{henry2000}, \cite{charbonneau2000}. This important discovery 
immediately fixed the ambiguity of the planet mass due to orbital inclination, 
proved that the source of radial velocity variation is an orbiting companion 
and not a stellar surface phenomenon and allowed for the first time to derive 
the two basic parameters (mass and radius) of an extrasolar planet.   

After this ground-braking discovery, over twenty different projects started to 
monitor large ($\sim 100$ square degrees) chunks of the sky and look for the 
same kind of events in millions of stars, without a priori radial velocity 
velocity monitoring. Interestingly, the first photometric discovery came not 
from these small-telescope projects but from the OGLE project, operating a 
$1.3$~m telescope and originally devoted to microlensing surveys. Unlike 
HD~209458, OGLE-TR-56 \cite{udalski2002} is a faint, $V=16.6$~mag star. 
This mounted rather strong constraints on the possible spectrographs used for 
the validation of the system. Nevertheless, in spite of its faintness, the 
system was successfully verified (see \cite{konacki2003}). It was soon realized 
that small telescopes are vital for the more massive discovery of additional 
systems that are considerably brighter.  

In 2004, TrES-1, the first transiting extrasolar planet (TEP) was discovered 
by using small aperture wide-field automated telescopes \cite{alonso2004}. 
The Trans-Atlantic Exoplanet Survey (TrES) network hosted the STARE telescope,  
the one that was instrumental in the discovery of the first transiting exoplanet, 
HD~209458. Although the discovery of TrES-1 was a good sign that small telescope 
projects are, indeed, capable of making significant discoveries, it was not 
until the fall of 2006 when these projects started to yield the rate of discovery 
hoped for. The relative long delay between conceiving the idea and the first 
science output is attributed to several factors. This includes the time needed 
for the technical implementation (please note that these telescopes are autonomous 
with sophisticated software making decisions on telescope operation, data 
acquisition and basic image reduction), development of new methods for massive 
signal search and, especially, solve the problem of filtering out colored noise, 
accessibility of proper instrumentation for precise followup observations with 
the proper amount of telescope time to get accurate light curve and radial 
velocity solutions. Last, but not least, these objects are rather rare, with a 
$\sim 0.4$\% (or lower) true occurrence rate (see, e.g., \cite{guo2016}) and 
an even lower (a factor of $10-30$ lower) observability rate. 
    
With the discovery of HAT-P-1 and WASP-1 \cite{bakos2007}, \cite{collier2007}  
the speed of the discoveries substantially increased, yielding by now over 
$200$ well-characterized HJ systems, covering a wide variety of stellar and planet 
parameters. Meanwhile, the long-planned space mission CoRoT was launched in 
2006 and started to deliver high-quality, uninterrupted photometric time series 
on thousands of stars, leading to the discovery of the first super-Earth object 
Corot-7b. Three years later, after many years of planning, the Kepler satellite 
was launched and a new era has begun with startling discoveries, such as Kepler-9, 
the first system with multiple transiting planets \cite{holman2010}, or Kepler-186f 
the first of the several systems hosting planets in the habitable zone 
\cite{quintana2014} , some of which might even be similar to Earth, such as 
Kepler-22b \cite{borucki2012}. The Kepler mission still continues, from 2014 under 
the name K2, showing the ingenuity of engineering and community support for the 
brilliant solution of the lost of the two reaction wheels by the time the main 
mission was completed. The K2 mission has proven to be quite successful. By 
visiting ecliptic fields, a broad spectrum of astrophysically interesting objects 
are visited, driven by researchers across the various disciplines of astronomy.   
 
The main science (discover and characterize planets outside the solar system) 
that drives the above ground- and space-based projects resulted in an increase 
in the quality and amount of the photometric data never seen before. It is 
obvious that this fundamental change has serious impact also on variable star 
research. In this review we focus on the ground-based wide-field surveys as 
the results of the variable star works on the space mission data are more widely 
known due to the large number of researchers working on these data. On the other 
hand, works on the variable star aspects of the ground-based surveys are less 
organized and sporadic. Nevertheless, as we will see, these surveys are both 
complementary and in many respects competitive to the space surveys. Therefore, 
it is very important to be aware of the results obtained so far and also of what 
these data could be used for in future works. 

%
%
\section{Wide-field photometric surveys - the overall importance}\label{sec:sec-2}
Unlike variable searches, targeting specific stars in earlier surveys (e.g., 
globular cluster studies, works on RR~Lyrae stars in microlensing survey 
data \cite{alcock1996}), current photometric projects, while looking for periodic 
transit signals, scan through the whole list of objects passing some minimum 
precision condition. Photometric surveys targeting extrasolar planets have to 
satisfy high standards in data quality (continuity and precision) and in volume 
(due to the rare occurrence and short duration of the event, even though it is 
periodic). This results in time series that surpass the quality of most of the 
photometric data gathered earlier and yields valuable source of a large diversity 
of various studies, not directly related to extrasolar planetary science. These 
science ``by-products'' may often be as important as those targeted by the 
original idea that has led to the initialization of the project.  

Many projects started after the first discovery of the transit of HD~209458 
in 2000. Some of them were not realized (e.g., Permanent All Sky Survey, or 
PASS\footnote{http://www.iac.es/proyecto/pass/}, \cite{deeg2004}), some of them 
were abandoned (e.g., Wise observatory Hungarian-made Automated Telescope, 
or WHAT\footnote{\url{http://wise-obs.tau.ac.il/~what/index.html}}, and some 
of them survived, expanded and became important/major contributors to the 
exoplanet inventory (e.g., Wide Angle Search for Planets, or WASP, 
\cite{pollacco2006}). 

In the design of the instrumentation of these surveys, one might consider 
the optimization of the system (including hardware and observational strategy) 
for the most effective survey of large number of stars. One of the issues is 
the size of the optics used. In an interesting paper by \cite{pepper2003}, 
the authors come up with a solution that employing a $2$~inch telescope, with 
a $4K\times4K$ CCD camera would be optimum. Indeed, this setting is very close 
to the ones used by most of the ``classical'' surveys that range from 
$4$~cm (KELT) to $\sim 20$~cm (e.g., HATSouth) with $10$~cm as the most often 
used and also, the combination of these under the same closure 
(Qatar Exoplanet Survey, or 
QES\footnote{\url{http://www.qatarexoplanet.org/}}, \cite{alsubai2016}). 

%
%
\begin{figure}[h]
\centering
\sidecaption
\includegraphics[width=5.5cm,clip]{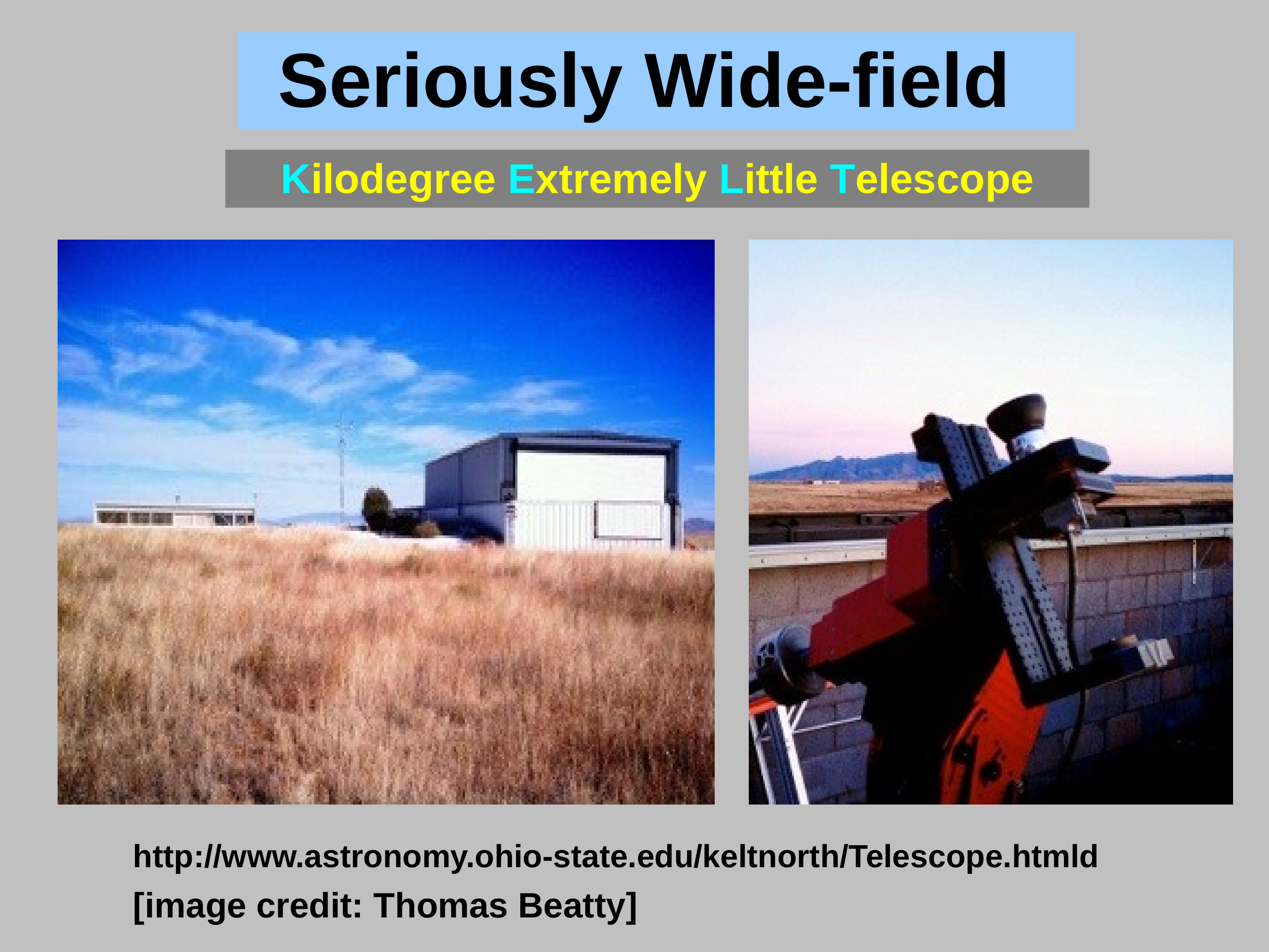}
\caption{The KELT-North building at the Winer Observatory (Arizona, USA), and 
the instrument hosted. The instrument is equipped with a $42$~mm diameter 
Mamiya f/1.9 telephoto lens, a $4K\times4K$ Apogee AP16E CCD camera put 
on a Paramount mount. The observations are fully robotic without the need 
for onsite supervision. The data are periodically transferred to the Ohio 
State University. 
}
\label{fig:KELT}
\end{figure}

For illustration, the somewhat unique setting for KELT-North is shown in 
Fig.~\ref{fig:KELT}. The instrument has a field of view of $26\times26$~square 
degrees, almost seven times as large as that of the instrument of the Kepler 
satellite. The southern station is hosted by the South African Astronomical 
Observatory  and constitutes also of a single instrument, the replica of the 
one at the northern station. Both instruments are fully robotic. They covered 
so far $\sim 70$\% of the sky, and supplied $4.5\times10^6$ light curves. As 
is today, they discovered $15$ TEPs\footnote{\url{http://exoplanet.eu}} and 
made possible visits to several important ``side topics'' 
(see Sect.~\ref{sec:sec-3}). 
 
\begin{figure}[h]
\centering
\sidecaption
\includegraphics[width=5.5cm,clip]{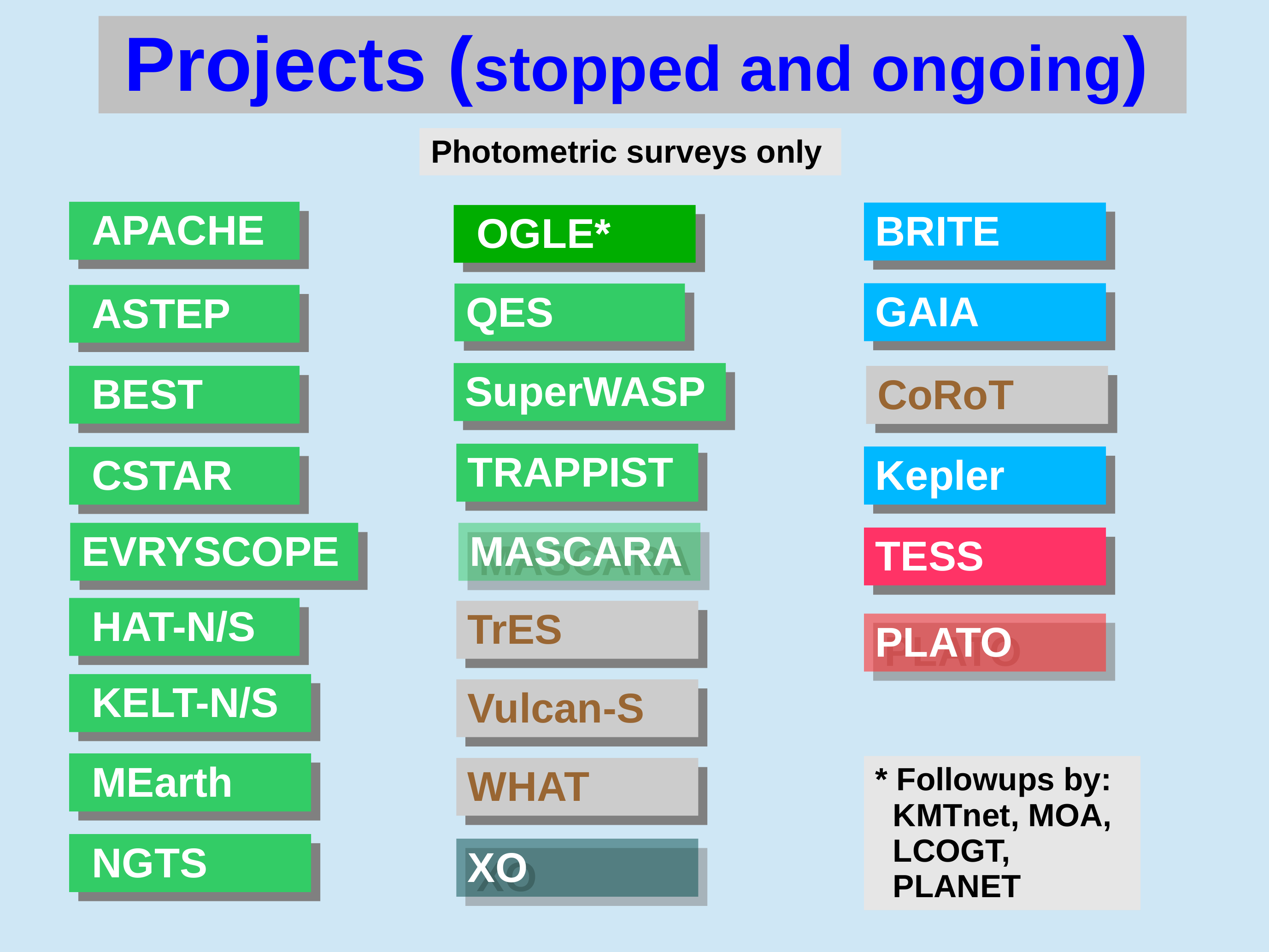}
\caption{Summary of the photometric survey programs with past/current/future 
contributions to exoplanet science. Different coloring/shading refer to the 
current status of these projects. The first two columns are for the ground-based 
projects, whereas the third column is for the space missions. Green and blue 
are for ongoing, gray for past, no longer running, red/orange for immediate/near 
future projects. The MASCARA project is at an early stage of operation 
\cite{talens2017}. The XO project has recently been revitalized \cite{crouzet2017}. 
The microlensing planet search conducted by the OGLE project tightly relies on 
the followup collaborative efforts with the participating projects/instruments 
shown in the footnote.}
\label{fig:projects}
\end{figure}

Fig.~\ref{fig:projects} summarizes the major photometric surveys contributed 
(or will contribute) to the discovery of TEPs. For a broader view, we included 
also related space mission\footnote{Gaia is not specified for planetary 
system search but will obviously yield a great contribution to it - mostly 
via the precise position measurements - but even through its low cadence 
photometry, once it is combined with dedicated ground-based followup 
observations (see \cite{dzigan2013}).}. It is impossible to give even a brief 
description of all these projects. Therefore, we focus on HAT-N/S and WASP, 
the two main surveys contributing to most of the TEPs discovered from the 
ground and describe some of the main parameters of these projects. However, 
it is important to note that several other projects also made very significant 
contributions. For example, 
MEarth\footnote{\url{https://www.cfa.harvard.edu/MEarth/Welcome.html}} 
(M dwarf stars in search of new Earth-like exoplanets, see \cite{irwin2014}) 
has important contribution to the variability survey of M dwarfs 
\cite{newton2016} and made the first discoveries of a transiting 
Super-Earth/Mini-Neptune \cite{charbonneau2009} and an Earth-sized planet 
from the ground \cite{berta2015}. Yet another, highly significant discovery 
comes from the 
TRAPPIST\footnote{\url{http://www.ati.ulg.ac.be/TRAPPIST/Trappist_main/Home.html}} 
(TRAnsiting Planets and PlanetesImals Small Telescope) project. The system consists 
of two $60$~cm robotic telescopes are located at La Silla Observatory (Chile) and 
at the Ouka\"imden Observatory (Marroco). In 2016 TRAPPIST made the first ground-based 
discovery \cite{gillon2016} of a multiplanetary system (consisting of three 
Earth-sized planets) around a nearby M8 dwarf. The proximity of the system makes 
it a prime candidate for deep study of the long-chased class of multiplanetary 
transiting planets around M dwarfs. And indeed, just before completing thus review, 
four additional planets were announced for this system, discovered by the Spitzer 
space telescope \cite{gillon2017}.  

%
%

%
%
\begin{figure}[h]
\centering
\sidecaption
\includegraphics[width=5.5cm,clip]{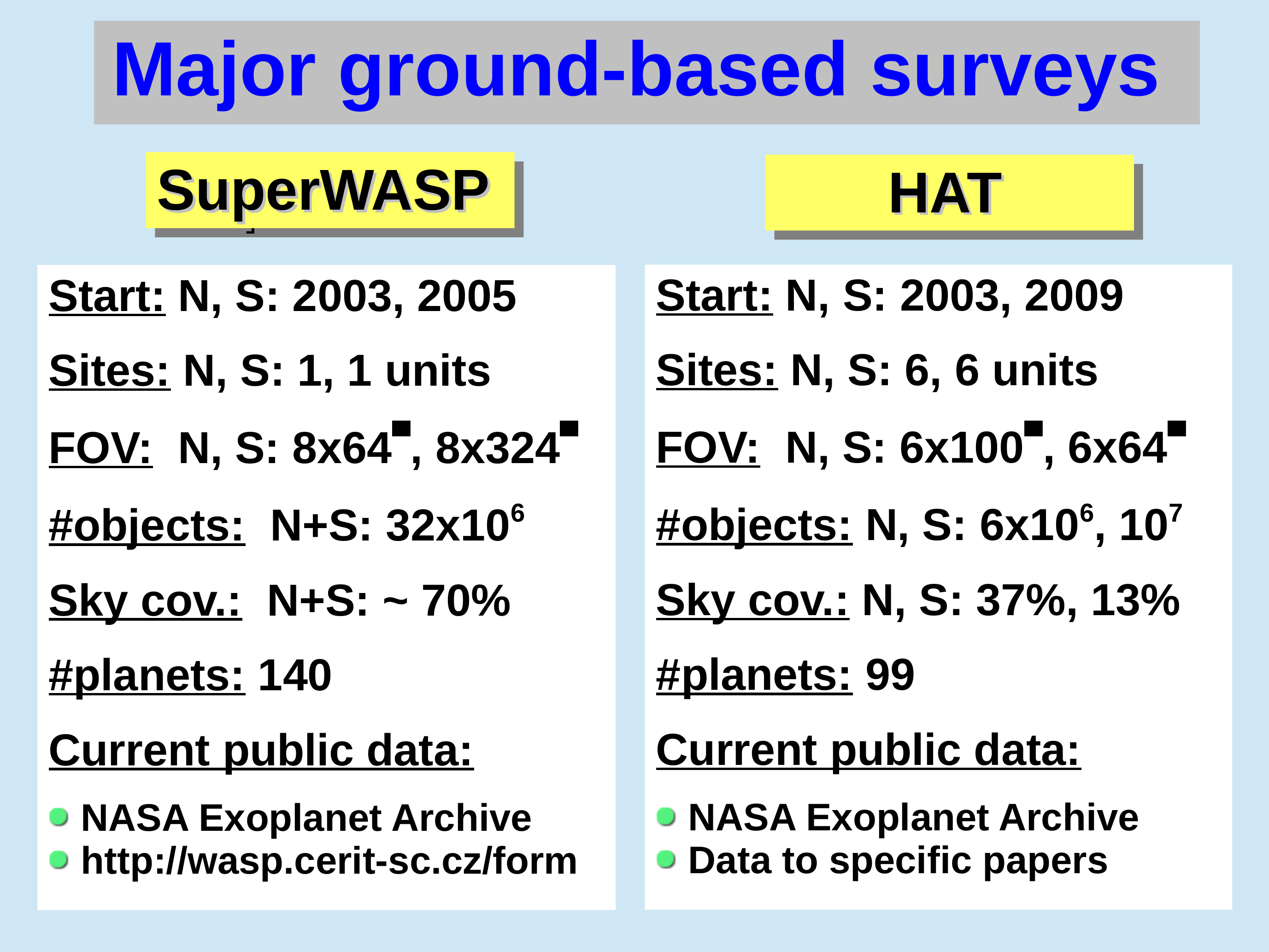}
\caption{Brief list of the main characteristics of the two major ground-based 
transiting extrasolar planet surveys. SuperWASP has recently switched to smaller 
diameter lenses at the Southern site. Earlier field of view was the same as that 
at the Northern site. Current sky coverages are based on private communications 
(SuperWASP) and on recently published papers (HAT-N/S, \cite{juncher2015}, 
\cite{bhatti2016}). Planet statistics are based on \url{http://exoplanet.eu} 
and contain a few commonly discovered systems.  
}
\label{fig:hat-wasp}
\end{figure}

Some of the basic characteristics of the two major survey programs are listed  
in Fig.~\ref{fig:hat-wasp}. SuperWASP\footnote{\url{http://www.superwasp.org}} 
consists of two observing units. One (SuperWASP-North) is in La Palma (Spain), 
the other one (SuperWASP-South) is at the South African Astronomical Observatory. 
Each unit consists 8 telephoto lenses (of $11$~cm and $7$~cm diameter) attached 
to $2K\times 2K$ CCD cameras enabling to take images on a very large chunk of the 
sky. The more detailed description of SuperWASP can be found in \cite{pollacco2006} 
and an update in \cite{smith2014}. 

The HAT project went through two major steps in the development. The survey on the 
Northern Hemisphere, HATNet\footnote{\url{http://hatnet.org/}}, started in 2004, 
and consists of two sites (the one at Mauna Kea Observatory/Havaii and the other 
at the Fred Lawrence Whipple Observatory/AZ) with altogether 6 observing units, 
each equipped with a $11$~cm diameter telephoto optics and a $4K\times 4K$ format 
CCD camera. The Southern counterpart (HATSouth\footnote{\url{http://hatsouth.org/}}) 
is a bigger undertaking, with observing units different in design than those of 
HATNet. Established in 2009, HATSouth consists of three sites: Las Campanas 
Observatory (Chile), Energy Stereoscopic System (HESS, Namibia) and Siding Spring 
Observatory (Australia). At each site there are two units with four $4K\times 4K$ 
CCD cameras attached to $18$~cm diameter astrographs. The detailed description 
of HATNet and HATSouth can be found in \cite{bakos2004} and \cite{bakos2013}, 
respectively. 

Both projects made (and are making) important discoveries in the field of extrasolar 
planets. With the spread in the longitude, careful data acquisition, in-depth 
treatment of the data and the longevity of the projects, these surveys are capable of 
detecting short period Hot Neptunes (e.g., HAT-P-11b \cite{bakos2010} and HAT-P-26b 
\cite{hartman2011b}), Warm Jupiters with periods longer than 10 days (e.g., HAT-P-15 
\cite{kovacs2010}, WASP-130 \cite{hellier2017}, HATS-17 \cite{brahm2016}) and Hot 
Jupiters with shallow transits around evolved stars (e.g., WASP-73 \cite{delrez2014}, 
HAT-P-50 \cite{hartman2015}). Both projects make the data used in their published 
papers available for the public. In addition, SuperWASP deposited all light curves 
gathered between 2004 and 2008 in a searchable format at the NASA/IPAC exoplanet 
site\footnote{Similar (albeit less extensive) depository exists for the KELT project 
and fractional data releases for TrES and XO -- 
see \url{http://exoplanetarchive.ipac.caltech.edu}}.

%
%
\section{Impact on data analysis - curing the red noise syndrome}\label{sec:sec-3}
As emphasized in Sect.~\ref{sec:sec-2}, wide-field surveys have to satisfy certain 
conditions in order to meet the original science goal. Because in the transit surveys 
all ``photometrically sound'' objects are searched for signals, any miniscule details 
count. Therefore, it is not surprising, that a considerable effort has been taken to 
achieve the highest data quality possible for a given instrumentation. In this section 
we briefly summarize the various methods that are intended to reach this goal. 

As follows from their very names, wide-field surveys cover a large part of the sky 
even if we consider only the single exposure images. Therefore, already the basic 
method of reduction (ensemble photometry, e.g., \cite{honeycutt1992}) should be 
modified \cite{bakos2004}, considering vignetting, differential reddening and 
refraction, change in the size and shape of the point spread function (PSF). 
Furthermore, even after taking all these into account, most of the stars will 
still contain some residual scatter due to uncured (or improperly considered) 
observational and environmental effects. Although a considerable filtering of 
these effects is possible with the aid of Differential Image Analysis (DIA, 
\cite{alcock1999}, or OIS \cite{alard1998}), even this, more in-depth treatment 
cannot cure all remaining systematics \cite{dekany2009}. Therefore, 
{\em post-processing} is an inevitable step in modern photometric surveys. 

%
%
\begin{table} 
\centering
\caption{Major red noise filtering methods developed during the past $12$ years}
\label{tab:tab-1}
\begin{tabular}{||l|l|l|l||}
\hline
 Acronym & Expanded name & Brief description$^*$ & Ref. \\
\hline\hline
{\bf TFA    } & Trend Filtering Algorithm   & Cotrending by the stars in the field using & \cite{kovacs2005} \\
{\bf        } &                             & simple least squares                        & \\
\hline
{\bf SysRem } & Systematics Removal         & Iterative step-by-step cotrending by the   & \cite{tamuz2005} \\
{\bf        } &                             & stars in the field, using PCA-related method        & \\
\hline
{\bf SARS   } & Simultaneous Additive and   & {\bf SysRem} extended to additive and       & \cite{ofir2010} \\
{\bf        } & Relative Sysrem             & multiplicative red noise                    & \\
\hline
{\bf EPD    } & External Parameter          & Using parameters (e.g., object pixel        & \cite{bakos2010} \\
{\bf        } & Decorrelation               & coordinates) external to the light curve    & \\
\hline
{\bf PDC    } & Presearch Data Conditioning & Cotrending by using Bayesian approach       & \cite{stumpe2012} \\ 
{\bf        } &                             & with PCA-selected template light curves     & \cite{smith2012} \\
\hline
{\bf MarPLE } & Marginalized Probability of & Using Bayesian method to find single transit& \cite{berta2012} \\ 
{\bf        } & a Lone Eclipse              & events and the period to combine them       & \\
\hline
{\bf TERRA  } & Transiting Exoearth Robust  & Cotrending by employing least squares fit,  & \cite{petigura2012} \\
{\bf        } & Reduction Algorithm         & PCA and template outlier selection          & \\
\hline
{\bf ARC    } & Astrophysically Robust      & Cotrending by using Bayesian approach       & \cite{roberts2013} \\
{\bf        } & Correction                  & with entropy criterion and PCA to select the& \\
{\bf        } &                             & essential contribution from the templates   & \\
\hline
{\bf SFF    } & Self Flat Fielding          & Systematics, due the roll correction of the & \cite{vanderburg2014} \\
{\bf        } &                             & Kepler (K2) spacecraft, are filtered out by & \\
{\bf        } &                             & using pixel positions as external parameters& \\
\hline
{\bf CPM    } & Casual Pixel Model          & Using pixel-level autoregressive model to   & \cite{wang2016} \\
{\bf        } &                             & predict systematics during transit          & \\
\hline
{\bf GPM    } & Gaussian Process Model$^{**}$ & Signal search by using Gaussian models    & \cite{aigrain2015} \\
{\bf        } &                             & for a simultaneous fit of the systematics   & \\
{\bf        } &                             & and the underlying signal                   & \\
\hline
{\bf FFF    } & Full-Fledged-Fit$^{**}$     & Periodic transit search by simultaneous fit & \cite{foreman2015} \\
{\bf        } &                             & for systematics and the signal              & \\
\hline
{\bf SIP    } & Systematics Insensitive     & FFF extended to sinusoidal signals          & \cite{angus2016} \\
{\bf        } & Periodogram                 &                                             & \\
\hline
{\bf DOHA   } & capital of Qatar            & Cotrending by best-correlating templates    & \cite{mislis2017} \\
{\bf        } &                             & on full and nightly time bases              & \\
\hline
\end{tabular}
\begin{flushleft}
$^*$ The description given may not fully reflect the full depth of the method. \\
$^{**}$ Our acronym attached to the method. 
\end{flushleft}
\end{table}

Differing in the details in the implementation, all these methods are based on the general 
idea that systematics are characterized by some combination of: a) the common perturbations 
in many objects in a given field, b) perturbations depending on external parameters, 
associated with each star, allowing peculiar (non-common) types of variability. Examples  
of type a) systematics are the nightly variation of the transparency, the trace of the 
improperly corrected track of an airplane or that of a cosmic ray. Type b) systematics 
include flux variability due to changes in the size and shape of the PSF, or to the 
spatial-dependent pixel sensitivity. The idea of systematics correction can be represented 
perhaps in the simplest way within the framework of TFA/EPD (see Table~\ref{tab:tab-1}). 
We approximate the observed flux variation $F(t)$ of our target of interest by the following 
expression 

%
%
\begin{eqnarray}
\label{eq:1}
\hat F(t) & = & F_0(t) + N(t) + \sum_{j=1}^{M}a_j X_j(t) + \sum_{i=1}^{K}b_i E_i(t) 
\hskip 2mm ,
\end{eqnarray}    

where $F_0(t)$ is the ``true'' signal (free from systematics and random - uncorrelated - 
noise), $N(t)$ is the instrumental/environmental white noise component, 
$\{X_j(t): j=1,2, ... ,M\}$, $\{E_i(t): i=1,2, ... ,K\}$ are, respectively, the $M$ 
template/cotrending time series of a representative group of objects in the field 
and the $K$ external parameters. The regression coefficients $\{a_j\}$ and $\{b_i\}$ 
are determined by some multiple regression method, in the simplest case by standard 
least squares. Since $F_0(t)$ is not known a priory, the standard approach in the 
case of signal search is to assume that systematics dominate the observed signal and 
set $F_0(t)$ equal to zero. The period search is performed on the time series obtained 
after the subtraction of the so-determined contribution of systematics. Once the 
period is found, the signal reconstruction phase follows, in which the complete 
signal model, represented by Eq.~(1) is solved. For transit signals this step is 
iterative, whereas for multiperiodic signals, representable by Fourier series, 
the solution can be obtained in a single step \cite{kovacs2008}.  

Please note that the philosophy of all systematics filtering methods is fundamentally 
different from the standard ensemble approach. In the latter we {\em do not fit} the 
target time series but simply use the time-dependent total (ensemble) flux to correct 
for the effects of atmospheric variations by dividing the instantaneous target fluxes 
by the corresponding ensemble fluxes. Under ideal circumstances, this method filters 
out only the atmospheric component of the variation, but leaves the intrinsic variation 
of the object intact. On the other hand, all systematics correction methods listed in 
Table~\ref{tab:tab-1} {\em distort} the signal at some degree, and the main 
purpose of many methods is to minimize this effect without jeopardizing the 
filter efficiency too much. This can be attempted by using only the ``essential'' 
cotrending time series (estimated by Principal Component Analysis, e.g., TERRA) 
and formulate the problem in the Bayesian framework to allow proper weightings 
of the different contributions to the systematics (e.g., PDC, ARC). One may also 
try to employ the full model (see Eq.~\ref{eq:1}) already in the period search phase, 
by assuming a certain type of signal with limited number of parameters. Surprisingly, 
this, highly time consuming method can be implemented by a proper grid-search algorithm 
that runs with an acceptable speed \cite{foreman2015}, \cite{angus2016}. Yet another 
way of running ``full-fledged'' period searches is to employ a Gaussian Process model 
that allows to weight the different contributions to the signal by using properly 
chosen correlation kernels \cite{aigrain2015}. MarPLE, the method developed for 
transit search on the sparse data of MEarth \cite{berta2012} is aimed also at 
conserving the signal shape, by using single nights to find single transits and 
combine these results within the Bayesian framework on the full dataset to find 
the period. 

Unfortunately, these works lack a thorough comparison with the more traditional 
approach, in which the data are filtered first for systematics, and then, the 
signal search is performed on these filtered data (containing considerably less 
systematics and the signal -- albeit the latter with various degree of distortion). 
In a subsequent study by \cite{kovacs2016} this comparison was performed on the 
subsets of the HATNet and K2 databases. They found that full-fledged methods may 
have limited applicability due to problems related to the proper disentangling 
of the signal and systematics in the observed signal. It seems that the traditional 
approach (filter first, then analyze), works better. 

%
%
\section{Selected highlights - the variable star aspects}\label{sec:sec-4}
In addition to the highly competitive nature of the field of extrasolar 
planets, ground-based wide-field surveys have been making a significant 
impact also on various fields of variable star research. So far, altogether 
over hundred papers have been published by these projects on objects 
(directly) unrelated to exoplanets. Several of them have received a 
considerable attention from the particular field (e.g., stellar rotations 
among Galactic field K$-$M stars \cite{hartman2011a} and in the open 
clusters Come Berenices \cite{collier2009} and in the Hyades and Praesepe 
\cite{delorme2011}). The fields covered by these ``side'' studies range 
from rotating asteroids \cite{parley2008} through classical pulsating stars 
\cite{skarka2014} to supernovae \cite{siverd2015}. It is impossible to 
summarize all these topics in the depth they deserve. Therefore, here 
we go through the main fields of study only briefly. Nevertheless, a few 
topics (reflecting mainly the prejudice of the author of this review) will 
be discussed somewhat deeper.  

\begin{figure}
\centering
\begin{minipage}[t]{.48\textwidth}
\includegraphics[width=\textwidth,clip]{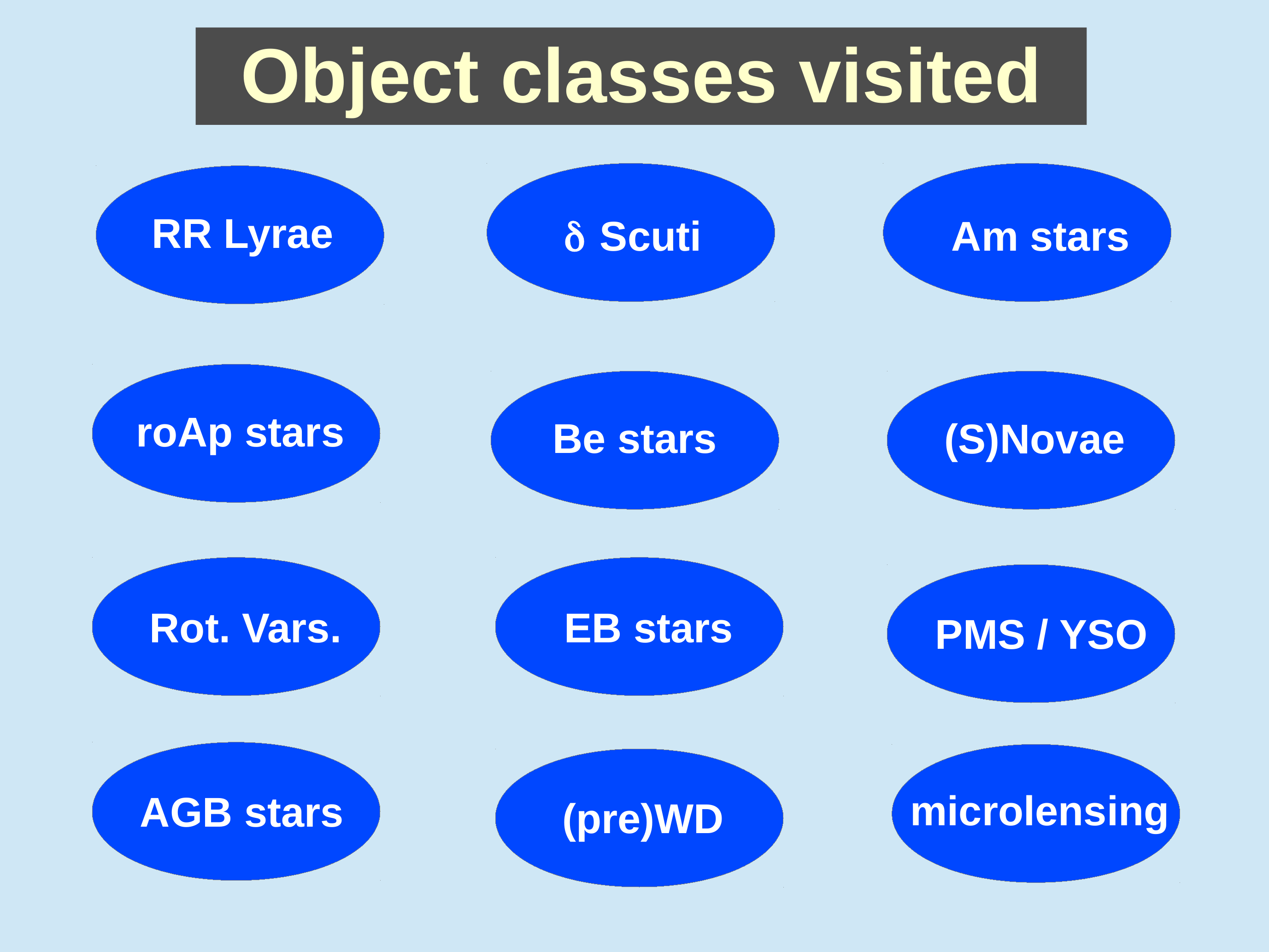}
  \caption{Classes of variable stars studied in various works using the data acquired 
  by ground-based wide-field surveys.}
  \label{fig:variable_topics}
\end{minipage}
\hfill
\begin{minipage}[t]{.48\textwidth}
\centering
\includegraphics[width=\textwidth,clip]{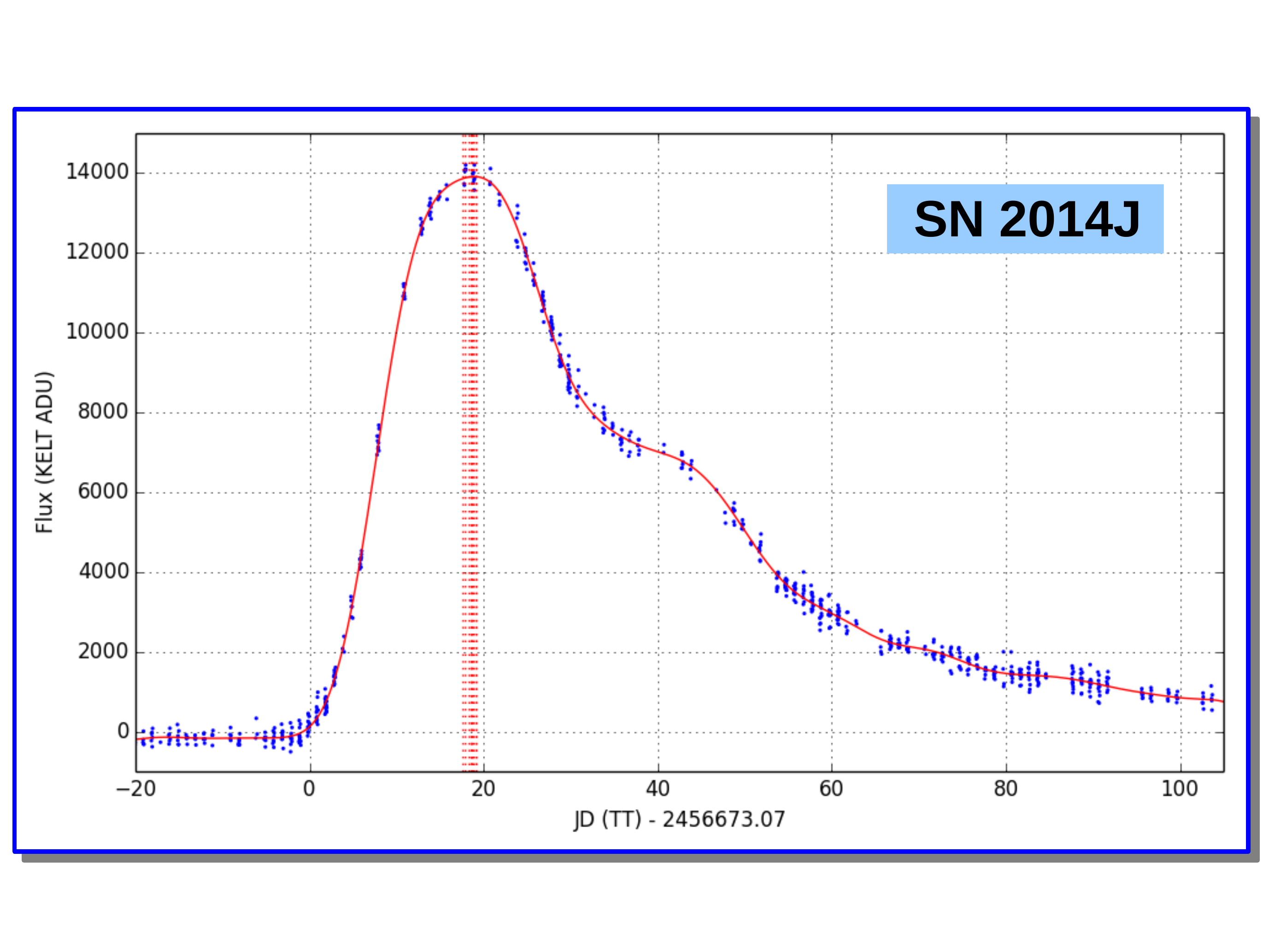}
  \caption{Detection of the pre- and post-outburst states of SN 2014J by using the 
  survey data of the KELT project \cite{siverd2015} (reproduced by the authors' 
  permission from \url{https://arxiv.org/}).}
  \label{fig:kelt_sn_2014j}
\end{minipage}
\end{figure}

Except for some special topics (e.g., Solar System bodies \cite{parley2008}), 
the main types of objects visited by the survey-related investigations are 
shown in Fig.~\ref{fig:variable_topics}. For curiosity, we also list 
``microlensing'', albeit this topic is limited to a single project only 
(for testing the ability of HATNet to detect microlensing events toward the 
Galactic Bulge \cite{nataf2009}).

%
%
%

\begin{itemize}
\item
{\em SN~2014J and eruptive stars:} 
By the very nature of the supernova explosions, they are usually discovered 
in the brightest phase, although some of them caught already during rising 
light (e.g., \cite{shappee2016}). On the other hand, the observation of the 
{\em pre-supernova} phase of SN~2014J in M82 \cite{siverd2015} was absolute 
unique until three similar events, back to 2011 and 2012, were disclosed 
in \cite{olling2015}, by using the archive of the Kepler satellite. The full 
light curve, as observed by the KELT project is shown in 
Fig.~\ref{fig:kelt_sn_2014j}. Analysis of the pre-burst state is very important 
for making distinction between the various progenitor scenarios leading to 
type Ia supernovae explosions. In the particular case of SN~2014J, other 
(e.g., HST, X-ray and radio) observations exclude the single degenerate$+$large 
companion model, but white dwarf mergers are allowed both by the above data 
and by the pre-burst light curve \cite{goobar2015}. 

Similarly exciting are the the discovery of the precursor of Nova Sco 2008 
\cite{tylenda2011} and the quite recent announcement of the nova candidate 
KIC 9832227 for an outburst in 2022 \cite{molnar2017}. In both cases the 
precursors are W~UMa binaries, and the prediction in the latter case relies 
on the observation of a similar steep decrease in the orbital period than 
for V1309 Sco, the binary progenitor of Nova Sco 2008. Both works utilize 
survey data, including those coming from wide-field projects (NSVS, ASAS, 
SuperWASP). 

For other eruptive phenomena, we draw attention for several studies performed 
by utilizing the SuperWASP survey. A handful of novae (recurrent and transient) 
were investigated by \cite{mcquillin2012}. Other works on eruptive variables 
can be found in \cite{byckling2009}, \cite{thomas2010}.
\item
{\em Pulsating sub-dwarf B stars:} 
SuperWASP pioneered the extension of variability search in the frequency 
regime close to or above the sampling rate used in the standard exoplanet 
survey mode. This has led to various exciting discoveries, including 
rapidly oscillating Ap stars (see later) and pulsating sdB stars (see 
Fig.~\ref{fig:J0902_freq_sp}). 
%

%
%
%
\begin{figure}
\centering
\begin{minipage}[t]{.48\textwidth}
\includegraphics[width=\textwidth,clip]{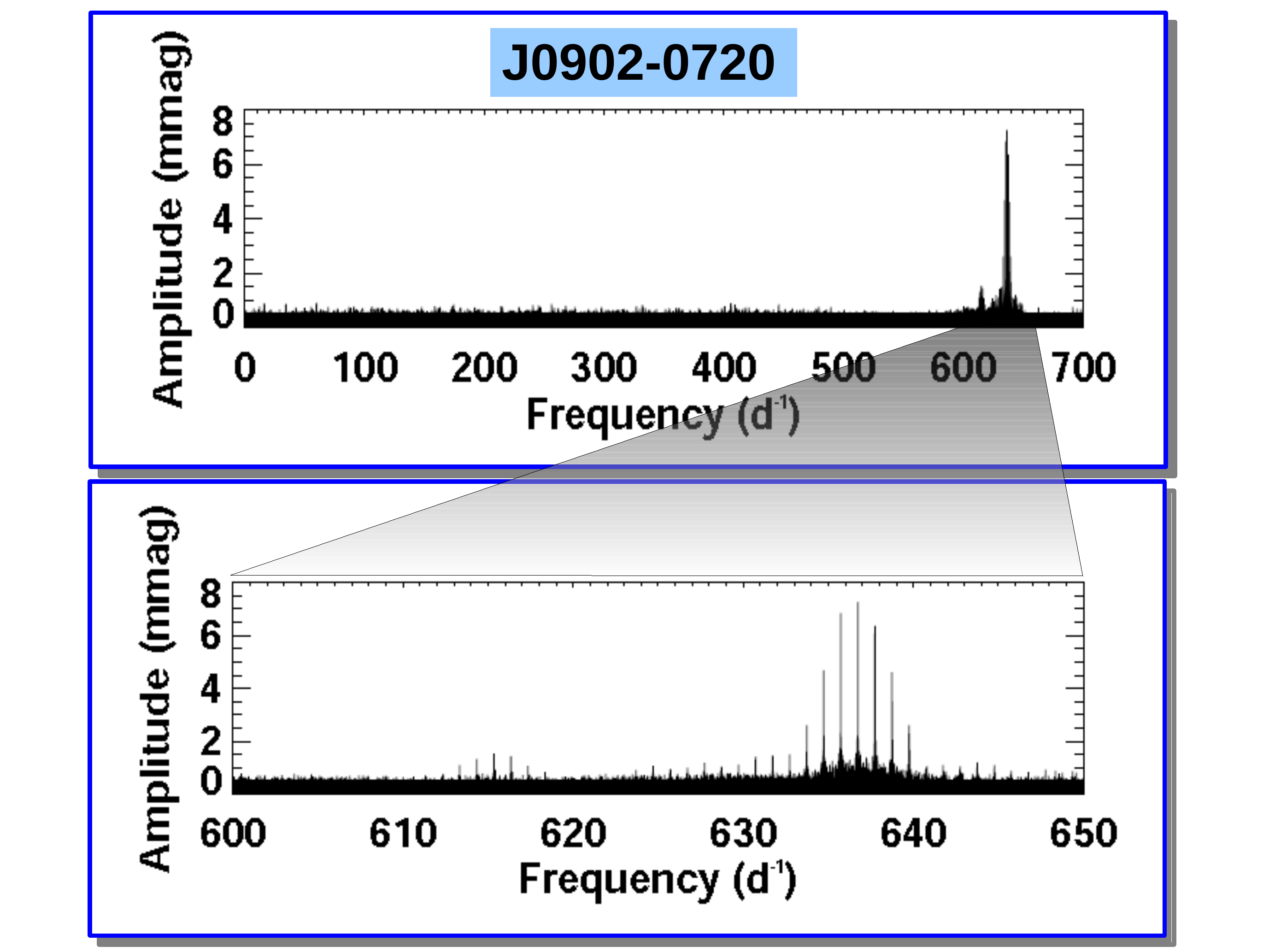}
  \caption{Frequency spectrum of the subdwarf B variable J0902 discovered by 
  \cite{holdsworth2017} using the survey data of the SuperWASP project. {\em Upper 
  panel:} full spectrum, {\em Lower panel:} zoomed in the high-power regime. Please 
  note the absence of low-frequency alias due to the lack of strong periodicity 
  in the data sampling. (Adapted by the authors' permission from 
  \url{https://arxiv.org/}.)}
  \label{fig:J0902_freq_sp}
\end{minipage}
\hfill
\begin{minipage}[t]{.48\textwidth}
\centering
\includegraphics[width=\textwidth,clip]{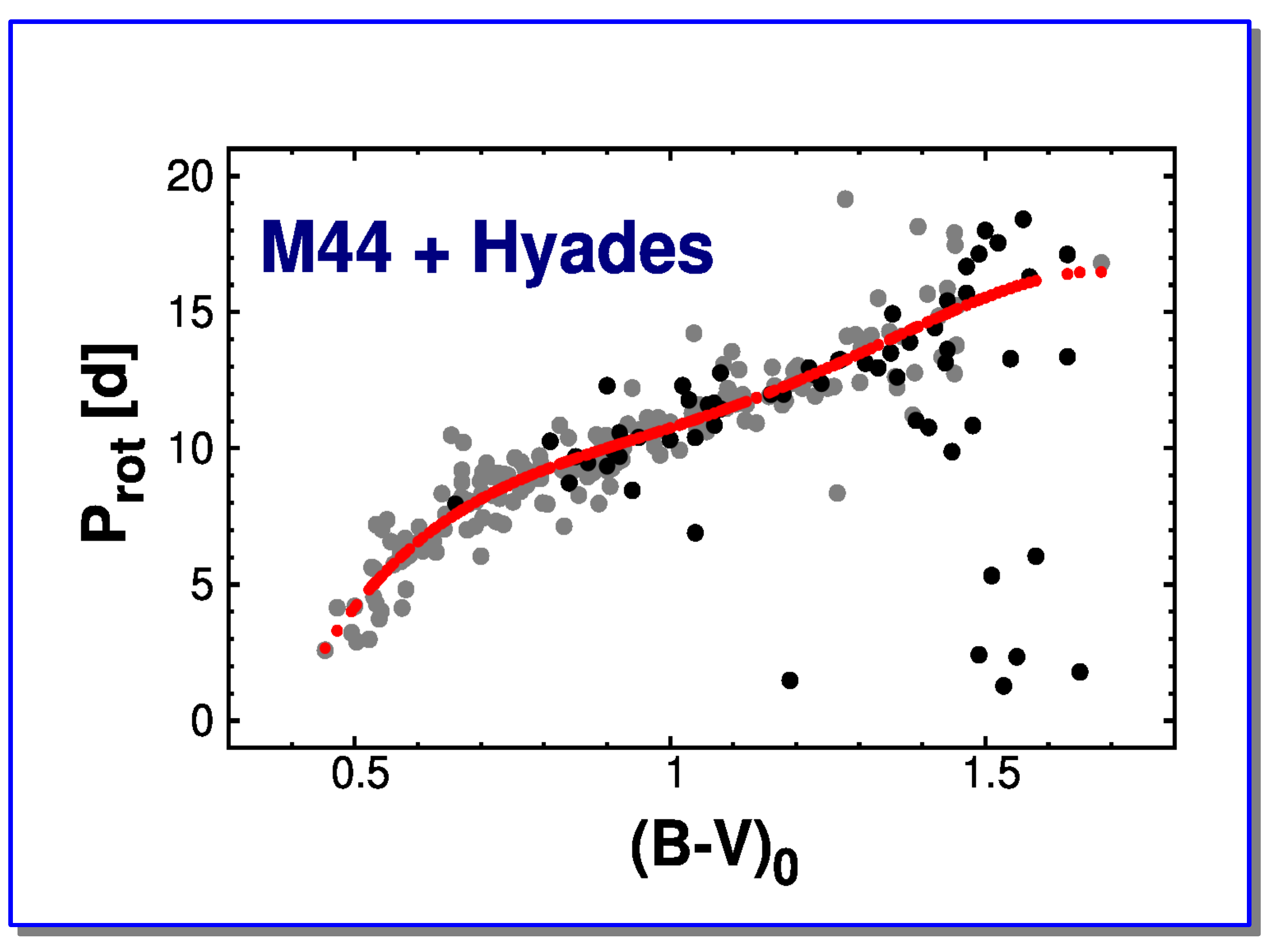}
  \caption{Color-Period diagram for the rotational variables of the open clusters 
  Praesepe and Hyades \cite{kovacs2015} (reproduced with permission \copyright~ESO). 
  Red dots show the fourth-order polynomial fit to the combined data of Praesepe 
  (gray points, HATNet) and the Hyades (black points, SuperWASP \cite{delorme2011}). 
  The two clusters have similar ages, leading to similar ridges for the rotationally 
  settled stars.}
  \label{fig:praesepe}
\end{minipage}
\end{figure}

\item
{\em Rotating stars in open clusters:} 
Stellar rotation can be exhibited through the photometric variation induced 
by the varying spot coverage of the visible hemisphere of the star. Prior 
to the wide-field surveys, specific targets in the field and open clusters 
were the subjects of individual projects, often facing with the difficulties 
detecting low-amplitude non-stationary signals on the time scale of few days 
to month. On the other hand, this kind of variability is in the comfort zone 
of the wide-field surveys. As already noted at the beginning of this section, 
several pioneering works have been made on open clusters and on the field by 
SuperWASP, KELT and HATNet. 
 
Fig.~\ref{fig:praesepe} shows the combined diagnostic diagram of stellar 
rotation for two famous clusters of the same-age. The method of gyrochronological 
age determination of field stars is based on these kinds of diagrams \cite{barnes2003} 
(see, however, current controversy on the applicability of this method on  
field stars: \cite{kovacs2015}, \cite{vansaders2016}, \cite{buzasi2016}.   
\item
{\em $\delta$~Scuti stars and variability of A-type stars:}
Stars on and close to the main sequence with spectral types A to F are know 
to exhibit low-amplitude, short period pulsations, often identified as p- 
or g-mode pulsations or some mixture of these. One of the unanswered 
questions concerning these variables is the dependence of the occurrence of 
the pulsations on the physical parameters. In spite of the theoretical 
expectations, only a fraction of the stars pulsate in the instability strip. 
Low sensitivity of the variability surveys is no longer an issue. Indeed, even 
with the exquisite accuracy of the Kepler space telescope, we find that only 
12\% of the stars in the instability strip are $\delta$~Scuti or $\gamma$~Dor 
(or mixed) variables (44\% are spotted or eclipsing variables, and the 
remaining 44\% are constant \cite{bradley2015} -- see also \cite{bowman2016} 
for a more extended survey). Most of the variables have amplitudes above $1$~mmag, 
i.e., within the reach of ground-based wide-field surveys. Indeed, many papers 
have been published on these stars, most of them utilizing the SuperWASP data. 

One of the most exciting findings was the discovery of the large occurrence of 
pulsators among Am stars (although rare occurrence of Am star pulsations was 
known from sparse earlier works -- e.g., \cite{kurtz1995}). From the examination 
of $1600$ Am stars, \cite{smalley2011} found $200$ variables exhibiting pulsations 
similar to those of ``normal'' $\delta$~Scuti or $\gamma$~Dor stars. The excitation 
mechanism of these stars could result from some delicate mixture of gravitational 
depletion of metals, rotation and turbulence (see the spectroscopic survey of 
Am stars by \cite{smalley2017}. Further examination of $1.5$ million stars in the 
F$-$B spectral range also in the SuperWASP database by \cite{holdsworth2014} has 
led to the discovery of additional Am stars and $10$ rapidly oscillating Ap 
stars. (A class of chemically peculiar stars, discovered unexpectedly by \cite{kurtz1982} 
in 1982.)   
\item
{\em Other diversities:} 
Unfortunately, we have space only to name few of the many exciting results: 
dimming episodes in young stellar objects \cite{rodriguez2013}, \cite{rodriguez2016a}; 
active main sequence B stars: \cite{labadie2016} ; 
RR Lyrae stars with Blazhko effects \cite{skarka2014}, \cite{sodor2012}; 
AGB stars: \cite{arnold2015}; 
eclipsing variables: quadruple (quintuple?) system \cite{lohr2013}, the longest 
eclipse ever measured \cite{rodriguez2016b} (see also \cite{lipunov2016}), EB close 
to the short period limit \cite{lohr2014}, etc.  
\end{itemize}

%
%
\section{Conclusions}\label{sec:sec-5}
We cannot conclude without emphasizing the great importance of the Northern 
Sky Variability Survey, NSVS (based on the observations of the Robotic Optical 
Transient Search Experiment, ROTSE-I, PI: {\em Carl Akerlof}) and the All Sky 
Automated Survey, ASAS (PI: {\em Grzegorz Pojmanski}) that proved the concept, 
the significance and the competitive scientific role of small, fully robotic 
instruments. We should also recall the indefeasible role of {\em Bohdan Paczynski}, 
who was a very strong supporter of small telescope variability surveys. 

%
%
\begin{figure}[h]
\centering
\sidecaption
\includegraphics[width=5.5cm,clip]{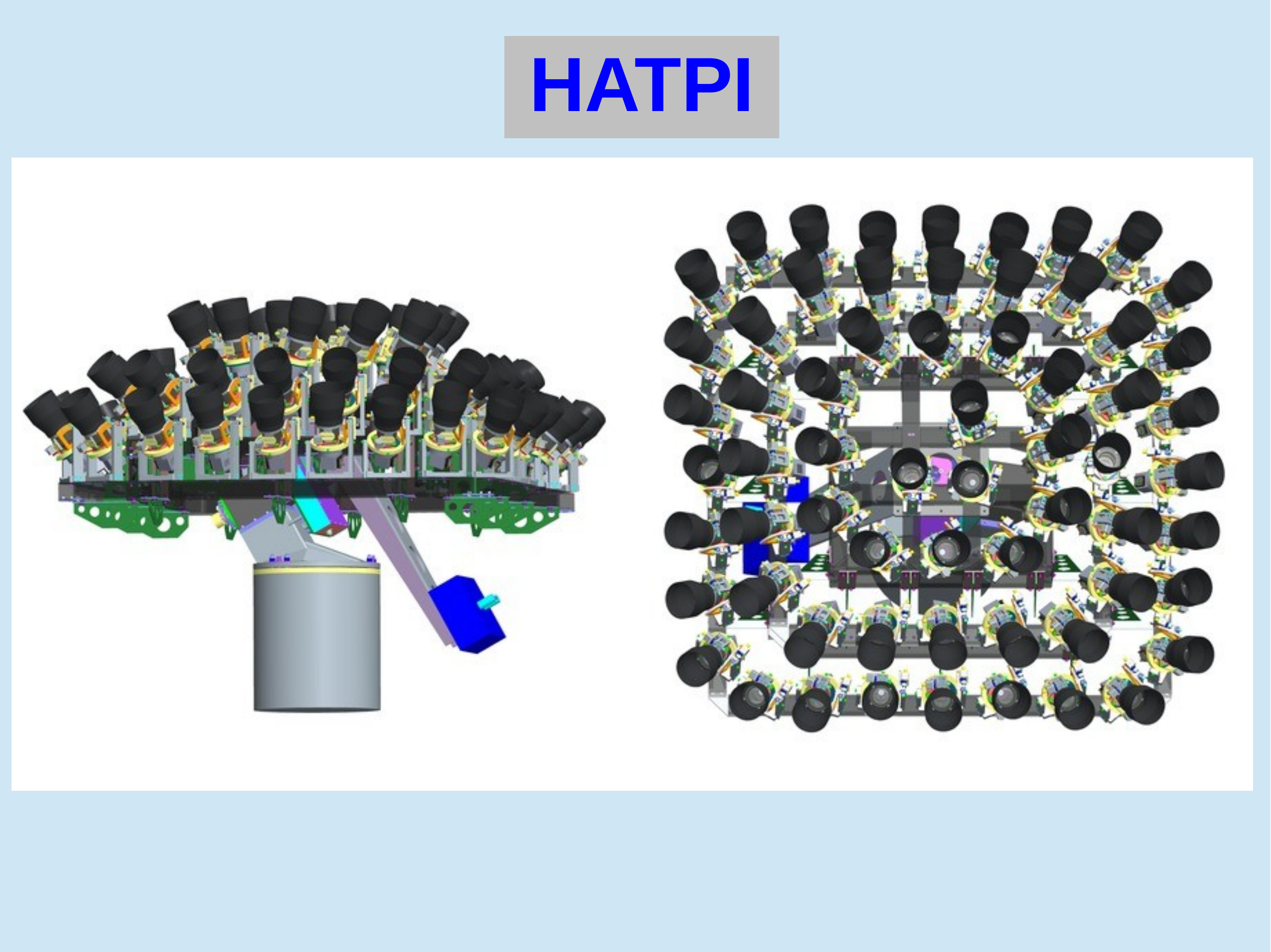}
\caption{Computer drawing of the HATPI platform, hosting 63 instrument-holder  
units (with optics, CCD and fine-pointer mechanics). The cameras will take images 
in every $30$ seconds virtually on the full visible sky at the Las Campanas 
Observatory. The project is in the active phase of development. When operational, 
it will be the biggest undertaking relative to other similar projects 
(EVRYSCOPE \url{http://evryscope.astro.unc.edu/}, 
FLY'S EYE  \url{https://flyseye.net/} and 
MASCARA    \url{http://mascara1.strw.leidenuniv.nl/}.}
\label{fig:hatpi}
\end{figure}

Today's wide field surveys grew up from the exciting and highly challenging 
idea of discovering extrasolar planetary systems by using the rare events 
of tiny dimmings of light when the planet moves across of the line of sight 
between the star and the observer. These projects have reached the level 
of efficiency when the discovery of this kind of events became standard. 
New methods of time series analysis have been developed, fertilizing other 
fields of research and helping in the development of the proper filtering 
of the data acquired by the various space missions. The data gathered by 
these projects are immense and we are only at the beginning of utilizing 
the millions of light curves observed over the past $15$ years. The examples 
shown in this summary suggest the wide range of applicability of these 
databases.   

Continuing and further developing ground-based wide field surveys will 
remain a significant goal in observational astronomy, in spite of highly 
competitive ground- and space-based projects (e.g., LSST, TESS). Full 
utilization of these data (including merging the different databases and 
the combination of them with other data from the ground and space) is still 
ahead. New projects are in progress (e.g., see Fig.~\ref{fig:hatpi}), 
aiming at more continuous sky monitoring with the goal of covering great 
variety of astrophysical phenomena from transiting planets to supernovae. 
We are sure that these assets will constitute an integral part of future 
variable star studies.  

%
%
\begin{acknowledgement} 
\noindent\vskip 0.2cm
\noindent {\em Acknowledgments}: Many thanks are due to the organizers of this 
conference to keep alive the reputation of the Los Alamos stellar pulsation 
meetings and for the cordial spirit throughout the conference. We gratefully 
acknowledge the information received on current project statuses from Don Pollacco, 
Richard West and Andrew Collier Cameron (SuperWASP), from G\'asp\'ar Bakos (HAT) 
and from Joshua Pepper (KELT). Prompt responses to my questions from Keivan Stassun 
and Daniel Holdsworth are appreciated. Thanks are due for the financial support 
received from the organizers and from the Directory Fund of the Konkoly Observatory.    
\end{acknowledgement}

\end{document}